\documentclass{article}
\usepackage[english]{babel}

\usepackage[a4paper,top=2cm,bottom=2cm,left=3cm,right=3cm,marginparwidth=1.75cm]{geometry}

\usepackage{graphicx}
\usepackage{amsmath}
\usepackage{tikz}
\usetikzlibrary{positioning}
\usepackage{caption}
\usepackage{authblk}
\usepackage{hyperref}
\usepackage{float}
\usepackage{url}
\usepackage{subcaption}
\usepackage{booktabs}

\hypersetup{
  colorlinks=true,
  linkcolor=blue,
  citecolor=blue,
  urlcolor=blue
}

\title{A Diagnostic Inverse-PINN Study of Identifiability in Reduced Black-Hole Spin Inference}
\author[1]{Stella Menziltsidou\thanks{Email: stmenzi@cs.duth.gr, ORCID: 0009-0009-2747-4373}}
\affil[1]{Department of Informatics, Democritus University of Thrace, Greece}

\begin{document}
\maketitle

\begin{abstract}
Black-hole spin is a fundamental parameter in relativistic astrophysics, influencing accretion efficiency, jet launching, and perturbative modes. This work investigates a hybrid physics-informed inverse framework for reduced black-hole spin inference. Spin recovery is formulated as an inverse problem constrained by a reduced scalar angular Teukolsky-like equation. A Physics-Informed Neural Network approximates the angular mode function, while the spin parameter is treated as a trainable physical quantity linked to the residual of the governing differential operator. The framework is evaluated under controlled synthetic and noise-contaminated angular-mode configurations across multiple reference spin values. The main result is diagnostic rather than confirmatory. The PINN reproduces angular-mode profiles qualitatively; however, the inferred spin values cluster near the upper part of the allowed interval rather than accurately recovering the full range of reference spins. This reveals a weak-identifiability regime in which accurate angular-profile reconstruction and low residual loss do not necessarily imply accurate spin recovery. The results highlight the diagnostic value of physics-informed constraints for reduced black-hole spin-inference experiments, while also showing the need for further identifiability analysis, loss reweighting, joint eigenvalue inference, full Kerr perturbation modeling, realistic uncertainty models, and validation against numerical solvers.
\end{abstract}


\section{Introduction}

Black-hole spin is one of the fundamental parameters of an astrophysical black hole, together with mass and, in principle, electric charge. It is commonly described by the dimensionless Kerr parameter
\begin{equation}
a_* = \frac{Jc}{GM^2},
\end{equation}
where $J$ is the angular momentum, $M$ is the black-hole mass, $G$ is the gravitational constant, and $c$ is the speed of light. The spin parameter affects several key physical processes, including the location of the innermost stable circular orbit, the radiative efficiency of accretion, the structure of relativistic jets, and the properties of perturbative modes around the black hole. For this reason, accurate and physically interpretable spin estimation remains an important problem in relativistic astrophysics.

Several observational and theoretical methods have been developed to constrain black-hole spin. The continuum-fitting method estimates spin by modeling the thermal emission of geometrically thin accretion disks. X-ray reflection spectroscopy infers spin from relativistically broadened reflection features, especially the iron K$\alpha$ line and the Compton reflection hump. Quasi-periodic oscillation models attempt to relate observed timing features to characteristic orbital or epicyclic frequencies near the compact object. Perturbation-theory-based approaches, including quasi-normal-mode analysis, connect spin to the oscillatory and damping properties of black-hole perturbations. These methods provide valuable constraints, but they also remain affected by modeling assumptions, calibration uncertainties, parameter degeneracies, source inclination, accretion-state dependence, and the availability and quality of observational data.

The Teukolsky formalism provides a central theoretical framework for studying perturbations of rotating black holes. It allows the dynamics of fields in the Kerr spacetime to be described through separable angular and radial equations. However, the full perturbation problem is mathematically and computationally demanding, especially when one aims to connect theoretical quantities with observational spin inference. Reduced angular formulations can therefore be useful as controlled spin-sensitive test problems, provided that their limitations are clearly stated. In particular, a scalar angular Teukolsky-like equation retains explicit dependence on the spin parameter and can be used to investigate how physics-informed learning methods may constrain spin-related quantities.

Machine-learning methods have increasingly been explored for astronomical parameter estimation, including applications to spectra, images, and time-series data. Purely data-driven neural networks can be flexible and scalable, but they usually require large labelled datasets and may behave as black-box estimators. They may also produce predictions that are not explicitly constrained by known physical equations. Physics-Informed Neural Networks (PINNs) address this limitation by embedding differential-equation residuals, boundary conditions, and other physical constraints directly into the training objective. This makes them particularly suitable for forward and inverse problems in which the available data are limited but the governing physical structure is known.

The present work proposes a hybrid physics-informed framework for black-hole spin estimation. The term ``hybrid'' is used in a methodological sense: the framework combines classical spin-estimation motivation, perturbation-theory structure, and physics-informed machine learning. Unlike a purely data-driven regression model, the proposed approach uses a reduced scalar angular Teukolsky-like operator as a spin-sensitive physical constraint. The spin parameter enters the differential operator and is therefore linked to the residual minimized during training.

A key point of the proposed formulation is that the spin parameter is not treated only as a prescribed configuration parameter. Instead, the spin-estimation problem is formulated as an inverse problem in which the angular mode function and the spin parameter are inferred consistently. A Physics-Informed Neural Network approximates the angular function, while the spin parameter is optimized as a trainable physical quantity. The total objective combines the differential-equation residual, boundary-condition penalties, angular-mode data consistency, normalization, and regularization terms. In this way, the estimated spin is constrained both by data agreement and by the underlying reduced physical equation.

This work also builds on the study associated with \cite{Menziltsidou2026EPJP}, which provides the related solver-level foundation for a simplified scalar angular Teukolsky-like equation. The present manuscript extends that direction by embedding the reduced solver into a broader inverse spin-estimation framework. Thus, the solver component is not used to test residual minimization for prescribed spin values; it is instead placed within a parameter-inference setting in which the spin itself becomes the quantity of interest.

The main contributions of this work are as follows:
\begin{enumerate}
    \item A hybrid inverse-PINN framework is introduced for black-hole spin estimation using a reduced scalar angular Teukolsky-like equation.
    \item The spin parameter is treated as an inferred physical quantity rather than only as a fixed input parameter.
    \item The loss function combines differential-equation residual minimization, boundary-condition enforcement, angular-mode data consistency, normalization, and regularization.
    \item The framework is evaluated under controlled synthetic and noise-contaminated angular-mode configurations across multiple reference spin values.
    \item Spin recovery is assessed using direct parameter-estimation metrics, including absolute error, relative error, mean absolute error, root mean squared error, and variability across random seeds.
\end{enumerate}

The remainder of the paper is organized as follows. Section~2 reviews classical black-hole spin-estimation methods, the Teukolsky formalism and physics-informed learning. Section~3 presents the proposed inverse physics-informed formulation. Section~4 describes the experimental design, including synthetic spin-recovery and noise-contaminated experiments. Section~5 reports the quantitative results. Section~6 discusses the astrophysical interpretation, limitations and relation to previous work. Section~7 concludes the paper and outlines future extensions toward full Kerr perturbations and observational spin-inference applications.

\section{Related Work}

\subsection{Classical Black-Hole Spin Estimation Methods}

Black-hole spin has traditionally been constrained through several observational and theoretical approaches. The continuum-fitting method estimates the spin parameter by modeling the thermal emission from a geometrically thin and optically thick accretion disk. In this framework, the inner edge of the disk is associated with the innermost stable circular orbit, which depends on the Kerr spin parameter. The method can provide strong constraints for stellar-mass black holes, but it requires independent estimates of the black-hole mass, distance, and disk inclination \cite{mcclintock2011measuring}.

X-ray reflection spectroscopy provides another widely used route for spin estimation. It models relativistically broadened reflection features, particularly the iron K$\alpha$ line and the Compton reflection hump, emitted from the inner accretion disk. Since the shape of these features depends on the relativistic geometry close to the black hole, reflection models can be used to infer the spin parameter. However, the method is affected by model degeneracies, disk-ionization assumptions, inclination effects and calibration uncertainties \cite{reynolds2014measuring, garcia2014improved}.

Quasi-periodic oscillation models attempt to connect observed timing features in X-ray binaries with characteristic frequencies near the compact object. These frequencies may be related to orbital, radial, or vertical epicyclic motions in the Kerr spacetime. Although QPOs provide potentially valuable spin-sensitive information, their physical interpretation remains model-dependent, and no universally accepted QPO model currently exists \cite{motta2015geometrical, stuchlik2016models}.

These classical methods provide important constraints, but they also illustrate the difficulty of black-hole spin estimation. Each method depends on modeling assumptions, observational quality, and source-specific parameters. This motivates complementary approaches that combine physical constraints with flexible inference methods.

\subsection{Black-Hole Perturbation Theory and Teukolsky-Type Equations}
Black-hole perturbation theory provides a theoretical route for connecting black-hole parameters with the behavior of fields and perturbative modes in curved spacetime. The Regge--Wheeler equation describes axial perturbations of non-rotating Schwarzschild black holes \cite{ReggeWheeler1957}, while the Zerilli equation describes the corresponding polar perturbations \cite{Zerilli1970}. For rotating Kerr black holes, the Teukolsky formalism provides the fundamental separable perturbation equation for fields of different spin weights in the Kerr geometry \cite{Teukolsky1973}.

The Teukolsky equation is central to the study of perturbations of rotating black holes and to the computation of quasi-normal modes. Quasi-normal-mode calculations provide a link between the black-hole mass, spin, and the complex oscillation frequencies of perturbations \cite{leaver1985analytic,berti2006gravitational}. In the gravitational-wave context, this connection underlies black-hole spectroscopy, where the ringdown signal can be used to constrain the properties of the remnant black hole.

The full Teukolsky problem is mathematically and computationally demanding. It involves spin-weighted fields, angular separation constants, complex quasi-normal-mode frequencies, and physically appropriate boundary conditions. For this reason, reduced scalar angular formulations can be useful as controlled spin-sensitive test problems. Such reduced equations do not replace the full Kerr perturbation problem, but they retain explicit dependence on the spin parameter and can therefore be used to investigate physics-informed spin inference in a simplified setting.

The present work uses a reduced scalar angular Teukolsky-like equation as the physical constraint inside an inverse learning framework. The goal is not to claim that the reduced equation is equivalent to the full Teukolsky master equation. Rather, the goal is to use its spin-dependent angular structure as a tractable physical constraint for developing and testing a hybrid spin-estimation methodology.

\subsection{Machine Learning and Physics-Informed Neural Networks}

Machine-learning methods have increasingly been used in astronomy and astrophysics for classification, parameter estimation, time-series analysis, and surrogate modeling. Purely data-driven neural networks can be highly flexible, but they generally require large labelled datasets and may behave as black-box estimators. In physical problems, this can be limiting because the model predictions are not necessarily constrained by the governing equations.

Physics-Informed Neural Networks (PINNs) address this limitation by embedding differential-equation residuals, boundary conditions, and other physical constraints directly into the training objective \cite{raissi2019physics}. Instead of learning only from labelled input-output pairs, a PINN is trained to satisfy a physical equation at collocation points while also fitting available data. This makes PINNs suitable for both forward problems, where the physical parameters are known, and inverse problems, where unknown parameters must be inferred from data.

In the context of inverse problems, PINNs can estimate unknown physical quantities by optimizing them jointly with the neural-network weights. This is particularly relevant for black-hole spin estimation because the spin parameter enters the differential operator itself. Therefore, the inferred spin can be constrained not only by data agreement but also by the residual of the governing equation. Uncertainty-aware extensions, including Bayesian PINNs and adversarial uncertainty quantification, further motivate the use of physics-informed models in noisy-data settings \cite{Yang2021BPINN,Yang2019AUQ}.

\subsection{Physics-Informed Learning for Black-Hole Perturbation Equations}

Recent studies have explored the use of PINNs and related physics-informed approaches for black-hole perturbation equations. Luna et al. used PINNs to compute quasi-normal modes of Kerr black holes through the Teukolsky equation, extracting complex frequencies and separation constants with accuracy typically below the percent level compared with accepted reference values \cite{luna2023solving}. Cornell et al. investigated supervised and unsupervised PINN approaches for solving Regge--Wheeler and Teukolsky equations, including higher overtone modes and multiple spin configurations \cite{cornell2024solving}.

The study \cite{Menziltsidou2026EPJP} is directly relevant to the present study. That work provides the solver-level foundation for a simplified scalar angular Teukolsky-like equation. The present manuscript extends that direction by embedding the reduced solver into a broader inverse spin-estimation framework. Also, the study \cite{Menziltsidou2026EPJP} establishes the reduced physics-informed solver component, while the present work develops the corresponding spin-inference formulation.

A forward PINN solver evaluates whether a neural network can satisfy a differential operator for prescribed physical parameters. The present study focuses instead on the inverse problem: estimating the black-hole spin parameter by combining angular-mode data consistency with the residual of a spin-dependent differential operator.

\subsection{Position of the Present Work}
The present work differs from purely observational spin-estimation methods, purely data-driven machine-learning regressors, and forward PINN solvers. Classical spin-estimation methods infer spin through observational signatures such as disk spectra, reflection features, timing frequencies, or perturbative modes. Data-driven machine-learning methods infer parameters from statistical correlations in training data. Forward PINN solvers typically assume the physical parameters are known and then approximate the solution of a differential equation.

The proposed framework combines these directions. It treats spin estimation as a physics-informed inverse problem. The angular mode function is approximated by a neural network, while the spin parameter is inferred as a trainable physical quantity. The inferred spin is constrained by both angular-mode data and the reduced Teukolsky-like operator. This provides a hybrid route toward physically constrained black-hole spin estimation.

\vspace{1em}
\begin{table}[h]
\centering
\caption{Comparison of Black Hole Spin Estimation Methods based on literature review \cite{mcclintock2011measuring,reynolds2014measuring,motta2015geometrical,baron2019machine,raissi2019physics,luna2023solving,cornell2024solving, Menziltsidou2026EPJP}.}
\label{tab:method_comparison}
\resizebox{\textwidth}{!}{
\begin{tabular}{|l|c|c|c|c|c|}
\hline
\textbf{Method} & \textbf{Physics-Based} & \textbf{Data-Driven} & \textbf{Interpretability} & \textbf{Scalability} & \textbf{Spin Sensitivity} \\
\hline
Continuum Fitting \cite{mcclintock2011measuring} & Yes & No & High & Low & Moderate \\
Reflection Spectroscopy \cite{reynolds2014measuring,garcia2014improved} & Yes & No & High & Moderate & High \\
QPO Models \cite{motta2015geometrical,stuchlik2016models} & Yes & No & Moderate & Low & High (source-dependent) \\
CNN/LSTM \cite{baron2019machine} & No & Yes & Low & High & Moderate \\
PINNs \cite{raissi2019physics,luna2023solving,cornell2024solving,Menziltsidou2026EPJP} 
& Yes & Yes & High & Moderate & High \\
\hline
\end{tabular}
}
\end{table}

\section{Methodology}

\subsection{Overview of the Hybrid Inverse Framework}

The proposed method formulates black-hole spin estimation as a physics-informed inverse problem. The central idea is that the black-hole spin parameter appears explicitly in the angular Teukolsky-like operator. Therefore, if angular-mode information is available, the spin parameter can be constrained by requiring the learned angular function to satisfy the corresponding spin-dependent differential equation.

The framework consists of three components. First, a reduced scalar angular Teukolsky-like equation is used as the physical constraint. Second, a neural network approximates the angular mode function. Third, the spin parameter is treated as an inferred trainable quantity and is optimized jointly with the neural-network weights. The resulting model combines data consistency with residual minimization, boundary-condition enforcement, normalization, and spin regularization.

\begin{figure}[htbp]
\centering
\includegraphics[width=\textwidth]{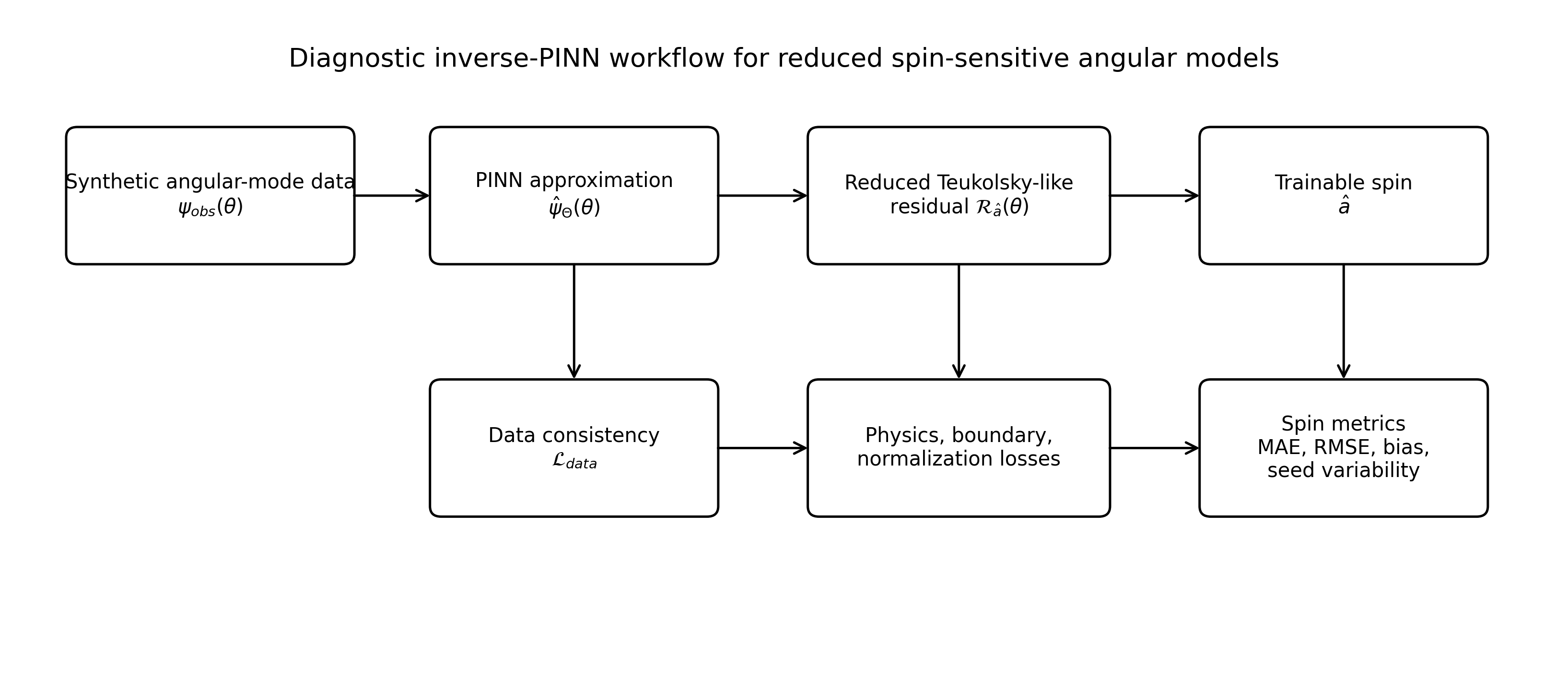}
\caption{Diagnostic inverse-PINN workflow for reduced spin-sensitive angular models. Synthetic angular-mode samples constrain the neural approximation, while the reduced Teukolsky-like residual and auxiliary loss terms provide physics-informed regularization for the trainable spin parameter.}
\label{fig:inverse_pinn_workflow}
\end{figure}

Figure~\ref{fig:inverse_pinn_workflow} summarizes the diagnostic inverse-PINN workflow used in this study. Synthetic angular-mode samples are provided to the neural approximation of the angular function, while the reduced Teukolsky-like residual, boundary and normalization conditions, and data-consistency terms jointly constrain the inferred spin parameter. The final output is not only the reconstructed angular profile, but also the inferred spin value and its associated error, bias, and seed-to-seed variability.

\subsection{Reduced Scalar Angular Teukolsky-Like Equation}

The physical constraint used in this work is a reduced scalar angular Teukolsky-like equation motivated by the angular sector of Kerr perturbation theory:
\begin{equation}
\frac{d}{d\theta}
\left(
\sin\theta \frac{d\psi}{d\theta}
\right)
+
\left[
\lambda
-
\frac{m^2}{\sin^2\theta}
-
a^2\omega^2\cos^2\theta
\right]\psi
=0.
\label{eq:reduced_teukolsky}
\end{equation}

Here, $\psi(\theta)$ is the angular mode function, $\theta$ is the polar angular coordinate, $a$ is the dimensionless black-hole spin parameter, $m$ is the azimuthal number, $\omega$ is the mode frequency, and $\lambda$ is the angular separation parameter. In the present implementation, $m$, $\omega$, and $\lambda$ are fixed to isolate the spin-recovery behaviour of the inverse formulation.

The reduced equation is not intended to reproduce the full Teukolsky master equation \cite{Teukolsky1973}. It is used as a controlled spin-sensitive operator that retains explicit dependence on the spin parameter. This allows the inverse PINN framework to be evaluated in a setting where the reference spin is known and the quality of spin recovery can be quantified directly.

\subsection{Inverse Spin-Estimation Formulation}
In the present inverse formulation, the spin parameter is unknown and must be inferred. The neural network approximates the angular mode function, while the spin parameter is optimized jointly with the network weights.

Let $\hat{\psi}_{\Theta}(\theta)$ denote the neural-network approximation of the angular mode function, where $\Theta$ represents the trainable network weights. The spin parameter is represented by a trainable scalar $\hat{a}$. The inverse problem is then formulated as the minimization of a composite physics-informed objective:
\begin{equation}
\mathcal{L}_{\mathrm{total}}
=
\mathcal{L}_{\mathrm{PDE}}
+
w_{\mathrm{BC}}\mathcal{L}_{\mathrm{BC}}
+
w_{\mathrm{data}}\mathcal{L}_{\mathrm{data}}
+
w_{\mathrm{norm}}\mathcal{L}_{\mathrm{norm}}
+
w_{\mathrm{reg}}\mathcal{L}_{\mathrm{reg}}.
\label{eq:total_loss}
\end{equation}

The inferred spin parameter is therefore not obtained from a purely statistical regression model. Instead, it is constrained by the angular-mode data and by the residual of the reduced spin-dependent differential operator.

\subsection{Physics-Informed Residual}

The PDE residual is obtained by applying the reduced Teukolsky-like operator to the neural-network prediction:
\begin{equation}
\mathcal{R}_{\hat{a}}(\theta)
=
\frac{d}{d\theta}
\left(
\sin\theta
\frac{d\hat{\psi}_{\Theta}}{d\theta}
\right)
+
\left[
\lambda
-
\frac{m^2}{\sin^2\theta}
-
\hat{a}^{\,2}\omega^2\cos^2\theta
\right]
\hat{\psi}_{\Theta}(\theta).
\label{eq:pde_residual}
\end{equation}

The physics-informed loss is then defined as
\begin{equation}
\mathcal{L}_{\mathrm{PDE}}
=
\frac{1}{N_f}
\sum_{i=1}^{N_f}
\left|
\mathcal{R}_{\hat{a}}(\theta_i)
\right|^2,
\label{eq:pde_loss}
\end{equation}
where $\{\theta_i\}_{i=1}^{N_f}$ are collocation points sampled from the angular domain. All derivatives are computed using automatic differentiation, following the general PINN framework for forward and inverse differential-equation problems \cite{raissi2019physics}.

\subsection{Boundary and Normalization Constraints}
Boundary-condition penalties are included to enforce the imposed angular endpoint behavior. In the present reduced scalar setting, homogeneous Dirichlet-type endpoint conditions are used:
\begin{equation}
\hat{\psi}_{\Theta}(0)=0,
\qquad
\hat{\psi}_{\Theta}(\pi)=0.
\end{equation}

The corresponding boundary loss is
\begin{equation}
\mathcal{L}_{\mathrm{BC}}
=
\left|
\hat{\psi}_{\Theta}(0)
\right|^2
+
\left|
\hat{\psi}_{\Theta}(\pi)
\right|^2.
\label{eq:bc_loss}
\end{equation}

A normalization condition is introduced to avoid convergence to the trivial zero solution:
\begin{equation}
\hat{\psi}_{\Theta}\left(\frac{\pi}{2}\right)=1.
\end{equation}

The normalization loss is
\begin{equation}
\mathcal{L}_{\mathrm{norm}}
=
\left|
\hat{\psi}_{\Theta}\left(\frac{\pi}{2}\right)-1
\right|^2.
\label{eq:norm_loss}
\end{equation}

\subsection{Angular-Mode Data Consistency}

To formulate spin estimation as an inverse problem, angular-mode samples are introduced through a data-consistency term. Let $\psi_{\mathrm{obs}}(\theta_i)$ denote the available angular-mode samples. These may correspond to clean synthetic samples or to noise-contaminated samples in the controlled experiments.

The data-consistency loss is defined as
\begin{equation}
\mathcal{L}_{\mathrm{data}}
=
\frac{1}{N_d}
\sum_{i=1}^{N_d}
\left|
\hat{\psi}_{\Theta}(\theta_i)
-
\psi_{\mathrm{obs}}(\theta_i)
\right|^2.
\label{eq:data_loss}
\end{equation}

This term links the learned angular function to the available data, while the PDE residual links the same function to the spin-dependent physical operator. The spin estimate is therefore constrained by both data agreement and physical consistency.

\subsection{Trainable Spin Parameter and Regularization}

The spin parameter is optimized jointly with the neural-network weights. To keep the inferred value within the physically relevant range, the trainable scalar is mapped to a bounded interval. For example, an unconstrained parameter $\alpha$ can be transformed as
\begin{equation}
\hat{a}
=
a_{\min}
+
(a_{\max}-a_{\min})\sigma(\alpha),
\label{eq:spin_mapping}
\end{equation}
where $\sigma(\cdot)$ is the logistic sigmoid function and $a_{\min}$ and $a_{\max}$ define the allowed spin range.

A regularization term is included to stabilize the inverse optimization:
\begin{equation}
\mathcal{L}_{\mathrm{reg}}
=
|\hat{a}|^2.
\label{eq:spin_reg}
\end{equation}

The final spin estimate is obtained by minimizing the full objective:
\begin{equation}
\hat{a}
=
\arg\min_a
\mathcal{L}_{\mathrm{total}}.
\label{eq:spin_estimate}
\end{equation}

\subsection{Possible Extension to Joint Eigenvalue Inference}
\label{sec:eigenvalue_extension}

In the present implementation, the angular separation parameter $\lambda$ is fixed in order to isolate the spin-recovery behavior of the reduced inverse formulation. However, in the full Teukolsky formalism, the angular separation constant is not generally an arbitrary fixed scalar independent of the spin and mode parameters. This motivates a natural extension in which $\lambda$ is inferred jointly with the spin parameter.

A joint inverse formulation may introduce a second trainable scalar $\hat{\lambda}$ and replace the residual in Eq.~\eqref{eq:pde_residual} by
\begin{equation}
\mathcal{R}_{\hat{a},\hat{\lambda}}(\theta)
=
\frac{d}{d\theta}
\left(
\sin\theta
\frac{d\hat{\psi}_{\Theta}}{d\theta}
\right)
+
\left[
\hat{\lambda}
-
\frac{m^2}{\sin^2\theta}
-
\hat{a}^{\,2}\omega^2\cos^2\theta
\right]
\hat{\psi}_{\Theta}(\theta).
\label{eq:joint_residual}
\end{equation}

The corresponding objective would optimize both $\hat{a}$ and $\hat{\lambda}$ jointly with the neural-network weights. This extension is expected to be important because fixing $\lambda$ may artificially restrict the solution family and may contribute to biased spin recovery. Joint eigenvalue inference would therefore provide a more physically faithful diagnostic step between the present reduced scalar model and a full Kerr perturbation treatment.

\section{Experimental Setup}

The experiments are designed to evaluate the proposed hybrid inverse physics-informed framework at two levels. First, a baseline forward PINN configuration is used as a diagnostic test to verify that the reduced scalar angular Teukolsky-like operator can be minimized under prescribed physical parameters. Second, the inverse formulation is evaluated by treating the spin parameter as an unknown trainable quantity and by measuring how accurately it can be recovered from angular-mode samples.

The baseline forward experiment is included only as a residual-convergence diagnostic. The main evaluation of the proposed spin-estimation framework is based on direct errors in the inferred spin parameter, including absolute error, relative error, mean absolute error, root mean squared error, and variability across random seeds.

\subsection{Baseline Forward PINN Configuration}

As an initial diagnostic, a baseline forward PINN was trained for a prescribed spin configuration. In this setting, the spin parameter was fixed to $a=0.7$, the azimuthal number to $m=0$, and the real-valued frequency parameter to $\omega=0.5$. The purpose of this experiment is not to perform spin inference, but to verify that the neural network can reduce the residual of the reduced scalar angular Teukolsky-like operator under controlled conditions.

The network was trained using the Adam optimizer with learning rate $10^{-3}$. Training was performed for 10,000 epochs using double-precision floating-point arithmetic. A total of 200 collocation points were sampled uniformly from the angular domain $\theta \in (0,\pi)$ and used to evaluate the physics-informed residual through automatic differentiation.

This baseline configuration should not be interpreted as a full complex quasi-normal-mode calculation. The frequency parameter is real-valued and the problem is restricted to the reduced scalar angular setting. Its role is to provide a controlled forward-solver reference before moving to the inverse spin-estimation experiments.

\begin{table}[h]
\centering
\caption{Baseline PINN training hyperparameters.}
\label{tab:hyperparams}
\begin{tabular}{|l|l|}
\hline
\textbf{Parameter} & \textbf{Value} \\
\hline
Input variable & $\theta \in (0,\pi)$ \\
Output variable & $\hat{\psi}(\theta)$ \\
Hidden layers & 4 \\
Neurons per hidden layer & 50 \\
Activation function & \texttt{tanh} \\
Optimizer & Adam \\
Learning rate & $10^{-3}$ \\
Epochs & 10,000 \\
Collocation points & 200 \\
Sampling strategy & Uniform in $\theta \in (0,\pi)$ \\
Boundary conditions & $\hat{\psi}(0)=\hat{\psi}(\pi)=0$ \\
Precision & float64 \\
\hline
\end{tabular}
\end{table}

\subsection{Baseline Loss-Convergence Diagnostic}

\begin{figure}[h]
\centering
\includegraphics[width=0.75\textwidth]{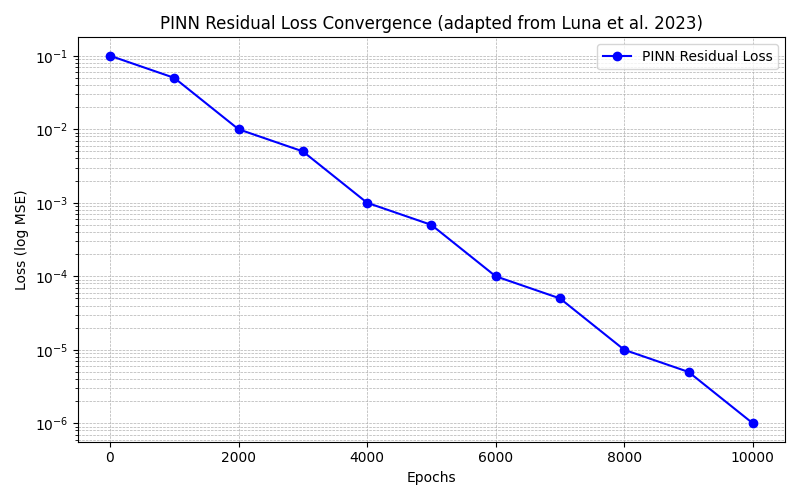}
\caption{Baseline forward PINN loss convergence for the reduced scalar angular Teukolsky-like equation with $a=0.7$, $m=0$, and $\omega=0.5$. The y-axis is shown in logarithmic scale.}
\label{fig:baseline_forward_loss}
\end{figure}

The baseline forward PINN loss convergence is reported as a diagnostic of residual minimization. The loss values at selected epochs are shown in Table~\ref{tab:baseline_forward_loss}, and the corresponding convergence curve is shown in Figure~\ref{fig:baseline_forward_loss}.

\begin{table}[h]
\centering
\caption{Baseline forward PINN loss convergence for the reduced scalar angular Teukolsky-like configuration with $a=0.7$, $m=0$, and $\omega=0.5$.}
\label{tab:baseline_forward_loss}
\begin{tabular}{|c|c|}
\hline
\textbf{Epoch} & \textbf{Total loss} \\
\hline
0     & $3.63 \times 10^{-1}$ \\
1000  & $6.18 \times 10^{-6}$ \\
2000  & $1.06 \times 10^{-6}$ \\
3000  & $6.07 \times 10^{-7}$ \\
4000  & $3.85 \times 10^{-7}$ \\
5000  & $2.10 \times 10^{-7}$ \\
6000  & $1.11 \times 10^{-7}$ \\
7000  & $7.05 \times 10^{-8}$ \\
8000  & $5.88 \times 10^{-8}$ \\
9000  & $5.37 \times 10^{-8}$ \\
\hline
\end{tabular}
\end{table}

As shown in Table~\ref{tab:baseline_forward_loss} and Figure~\ref{fig:baseline_forward_loss}, the total loss decreases from $3.63\times10^{-1}$ at initialization to $5.37\times10^{-8}$ near epoch 9000. This indicates that the network can reduce the residual of the reduced angular operator under the imposed boundary conditions.

Small oscillations in the later stages of training may be associated with optimizer dynamics, sensitivity of the PDE residual, and the interaction between residual and boundary-condition terms. This result is used as a forward-solver diagnostic only. It is not used as the primary evidence for spin-estimation performance, which is instead evaluated through direct errors in the inferred spin parameter.

\subsection{Inverse Spin-Recovery Experiments}

The main experiments evaluate the inverse spin-estimation formulation. In these experiments, the spin parameter is not prescribed during training as a fixed input. Instead, it is represented by a trainable scalar $\hat{a}$ and optimized jointly with the neural-network weights.

Synthetic angular-mode samples are generated for reference spin values
\[
a_{\mathrm{true}} \in \{0.2,0.3,0.5,0.7,0.9\}.
\]
For each reference spin value, the inverse PINN is initialized from the same starting value $a_{\mathrm{init}}=0.5$ and trained to recover the corresponding spin parameter. The inferred spin value $\hat{a}$ is then compared with the known reference value $a_{\mathrm{true}}$.

The purpose of this experiment is to test whether the spin-sensitive differential operator, together with angular-mode data consistency, provides sufficient information for recovering the spin parameter in the reduced scalar angular setting.

\subsection{Planned Loss-Ablation and Identifiability Diagnostics}
\label{sec:planned_ablation_diagnostics}

The systematic loss-ablation study is treated as a future diagnostic extension rather than as a completed benchmark in the present manuscript. Such an analysis should compare the full inverse-PINN objective against variants without spin regularization, with increased data-consistency weight, with increased physics-residual weight, and with adaptive loss reweighting. This would help determine whether the observed clustering of the inferred spin parameter is mainly caused by weak identifiability of the reduced angular operator or by the optimization geometry of the composite loss.

In the present work, the main completed diagnostic is the comparison between forward residual convergence and inverse spin recovery. The observed spin clustering motivates additional identifiability tests, including fixed-spin loss-profile scans, sensitivity analysis of the residual with respect to the spin parameter, and joint inference of the angular separation parameter.

\subsection{Noise-Contaminated Spin-Recovery Experiments}

To evaluate robustness, Gaussian perturbations are added to the synthetic angular-mode samples. The noisy samples are defined as
\begin{equation}
\psi_{\mathrm{obs}}(\theta_i)
=
\psi_{\mathrm{ref}}(\theta_i)
+
\epsilon_i,
\qquad
\epsilon_i \sim \mathcal{N}(0,\sigma^2),
\end{equation}
where $\psi_{\mathrm{ref}}(\theta_i)$ denotes the clean reference angular-mode profile and $\sigma$ controls the noise amplitude.

The tested noise levels are
\[
\sigma \in \{0.00,0.01,0.03,0.05,0.10\}.
\]
For each noise level, the inverse spin-recovery experiment is repeated across multiple random seeds. This allows the variability of the inferred spin parameter to be quantified and avoids relying on a single stochastic training run.

The Gaussian noise model is used as a controlled perturbation test. It is not intended to reproduce the full complexity of real observational uncertainties, such as detector response, calibration errors, inclination degeneracies, accretion-state dependence, or source-specific systematics.

\subsection{Additional Controlled Uncertainty Scenarios}
\label{sec:additional_uncertainty}

Although Gaussian perturbations provide a useful first robustness test, they do not represent the full structure of observational uncertainty in black-hole spin measurements. Therefore, the present diagnostic framework can be extended to additional controlled uncertainty scenarios that remain synthetic but are closer to realistic sources of error.

First, heteroscedastic perturbations can be introduced by allowing the noise amplitude to vary with angular coordinate,
\begin{equation}
\epsilon_i \sim \mathcal{N}(0,\sigma^2(\theta_i)).
\end{equation}
This tests whether the inferred spin is sensitive to non-uniform uncertainty across the angular domain.

Second, systematic calibration-like distortions can be applied through a multiplicative perturbation,
\begin{equation}
\psi_{\mathrm{obs}}(\theta_i)
=
(1+\delta)\psi_{\mathrm{ref}}(\theta_i)
+
\epsilon_i,
\end{equation}
where $\delta$ represents a controlled systematic offset.

Third, incomplete angular coverage can be simulated by removing samples from selected angular intervals. This is useful because real inference problems often involve incomplete or unevenly informative data. These additional tests would help distinguish robustness to random perturbations from robustness to systematic modeling effects.

\subsection{Evaluation Metrics}

The primary evaluation quantity is the accuracy of the inferred spin parameter. For each experiment, the absolute spin error is defined as
\begin{equation}
E_{\mathrm{abs}}
=
|\hat{a}-a_{\mathrm{true}}|,
\end{equation}
and the relative spin error is defined as
\begin{equation}
E_{\mathrm{rel}}
=
\frac{|\hat{a}-a_{\mathrm{true}}|}
{|a_{\mathrm{true}}|}.
\end{equation}

Across multiple spin values, noise levels, and random seeds, the mean absolute error is computed as
\begin{equation}
\mathrm{MAE}
=
\frac{1}{N}
\sum_{i=1}^{N}
|\hat{a}_i-a_{\mathrm{true},i}|,
\end{equation}
and the root mean squared error is computed as
\begin{equation}
\mathrm{RMSE}
=
\sqrt{
\frac{1}{N}
\sum_{i=1}^{N}
(\hat{a}_i-a_{\mathrm{true},i})^2
}.
\end{equation}

The standard deviation of $\hat{a}$ across random seeds is also reported in the noise-contaminated experiments. This provides a measure of the stability of the inverse optimization under different noise realizations and network initializations.

Residual loss is monitored as a diagnostic of physical consistency, but it is not treated as the main performance metric. The main performance criterion is the ability of the framework to recover the spin parameter itself.

\subsection{Reproducibility and Implementation}

All experiments are implemented in Python using automatic differentiation for the evaluation of the physics-informed residual. Double-precision floating-point arithmetic is used throughout the training procedure. The Adam optimizer is used for both the neural-network weights and the trainable spin parameter.

For reproducibility, the experiments store the inferred spin values, loss histories, error metrics, and generated figures. The final implementation includes CSV files for the spin-recovery results, noise-robustness statistics, and training histories, together with scalable PDF figures for journal-quality visualization.

The code and supporting material are made available in a public repository, including the scripts required to reproduce the forward loss-convergence diagnostic, the inverse spin-recovery experiments, and the noise-contaminated robustness analysis.

\section{Benchmarking and Synthetic Spin-Recovery Evaluation}

This section evaluates the proposed inverse physics-informed framework against controlled synthetic reference models. The purpose of the benchmark is to measure whether the method can recover the spin parameter from angular-mode information when the reference spin is known. The benchmark is therefore designed as a parameter-recovery test, rather than as a qualitative comparison between methods.

The evaluation is performed in three stages. First, clean synthetic angular-mode samples are used to test spin recovery in the absence of perturbations. Second, Gaussian noise is added to the synthetic angular-mode samples to evaluate robustness. Third, baseline synthetic models are described as methodological comparators, although quantitative baseline scores are not reported in the supplied result files. The primary benchmark quantities are the estimated spin $\hat{a}$, absolute spin error, relative spin error, mean absolute error, root mean squared error, and variability across random seeds.

\subsection{Synthetic Reference Models}

Synthetic reference models are generated for prescribed spin values
\[
a_{\mathrm{true}} \in \{0.2,0.3,0.5,0.7,0.9\}.
\]
For each value of $a_{\mathrm{true}}$, a clean angular-mode reference profile $\psi_{\mathrm{ref}}(\theta)$ is generated under the reduced scalar angular Teukolsky-like formulation. These reference profiles provide controlled angular-mode samples for which the true spin parameter is known.

The synthetic setting is used because it allows the spin-recovery error to be measured directly. This is important for benchmarking: unlike residual loss, which measures the degree to which the learned function satisfies the imposed differential operator, spin-recovery error measures the accuracy of the inferred physical parameter itself.

The synthetic benchmark does not claim to reproduce the full observational complexity of black-hole spin measurements. Instead, it provides a controlled environment for evaluating whether the proposed inverse physics-informed formulation can recover spin-sensitive information from angular-mode data.

\subsection{Baseline Synthetic Models for Comparison}

To contextualize the role of the physics-informed constraint, simplified synthetic baseline models are described as methodological comparators. These baselines are not intended to reproduce the full complexity of observational spin-estimation methods such as continuum fitting or X-ray reflection spectroscopy. Instead, they define possible reference points for future controlled numerical comparisons within the same synthetic angular-mode setting.

The following models are considered:

\begin{enumerate}
    \item \textbf{Data-only neural regression}: a neural model trained only to fit the angular-mode samples, without the Teukolsky-like residual term.
    \item \textbf{Least-squares synthetic fit}: a direct fitting baseline that minimizes the mismatch between the predicted and reference angular-mode samples over a grid or parameterized family of spin values.
    \item \textbf{Forward PINN diagnostic}: a physics-informed solver with prescribed spin, used only to verify residual minimization and not to infer the spin parameter.
    \item \textbf{Proposed inverse PINN}: the full model in which the spin parameter is trainable and constrained jointly by data consistency, boundary conditions, normalization, regularization, and the reduced Teukolsky-like residual.
\end{enumerate}

In the present manuscript, these baseline models are retained as methodological comparators rather than as completed quantitative benchmarks. This choice reflects the diagnostic aim of the study: the goal is not to claim numerical superiority over all possible synthetic baselines, but to evaluate whether the proposed inverse-PINN formulation can identify spin from reduced angular-mode information. A full quantitative comparison with data-only neural regression and least-squares synthetic fitting remains an important direction for future work.

\subsection{Clean Synthetic Spin-Recovery Benchmark}

The first benchmark evaluates spin recovery under clean synthetic conditions. For each reference spin $a_{\mathrm{true}}$, the inverse PINN is initialized from the same starting value $a_{\mathrm{init}}=0.5$ and trained to infer $\hat{a}$. The inferred value is then compared with the known reference spin.

\begin{table}[htbp]
\centering
\caption{Clean synthetic spin-recovery benchmark. The reported inferred spin is the mean value across five random seeds; the uncertainty term is the standard deviation across seeds. Absolute and relative errors are computed with respect to the corresponding reference spin.}
\label{tab:clean_spin_benchmark}
\begin{tabular}{ccccc}
\hline
$a_{\mathrm{true}}$ & $a_{\mathrm{init}}$ & $\hat{a}$ & Absolute error & Relative error \\
\hline
0.20 & 0.50 & $0.982425 \pm 0.000270$ & 0.782425 & 3.912127 \\
0.30 & 0.50 & $0.982397 \pm 0.000279$ & 0.682397 & 2.274656 \\
0.50 & 0.50 & $0.982436 \pm 0.000272$ & 0.482436 & 0.964871 \\
0.70 & 0.50 & $0.982533 \pm 0.000251$ & 0.282533 & 0.403618 \\
0.90 & 0.50 & $0.982495 \pm 0.000262$ & 0.082495 & 0.091661 \\
\hline
\end{tabular}
\end{table}

The clean benchmark tests whether the reduced spin-sensitive operator contains sufficient information for parameter recovery when the angular-mode samples are not perturbed by noise. This experiment is a necessary baseline before evaluating robustness under noisy synthetic conditions.

\begin{figure}[H]
\centering
\begin{subfigure}{0.4\textwidth}
\centering
\includegraphics[width=\textwidth]{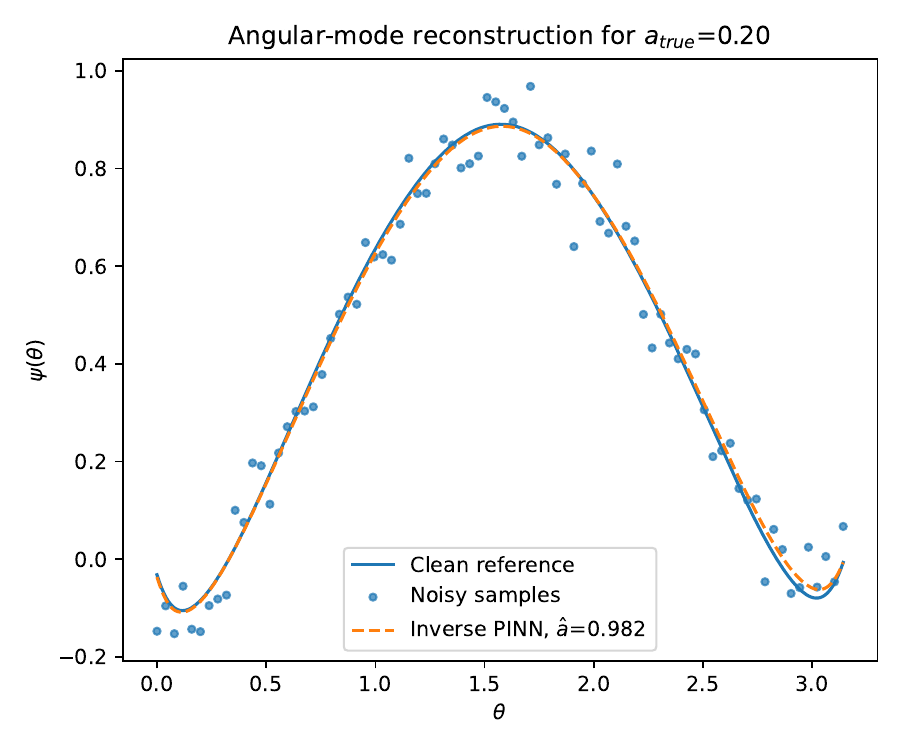}
\caption{$a_{\mathrm{true}}=0.20$}
\end{subfigure}
\hfill
\begin{subfigure}{0.4\textwidth}
\centering
\includegraphics[width=\textwidth]{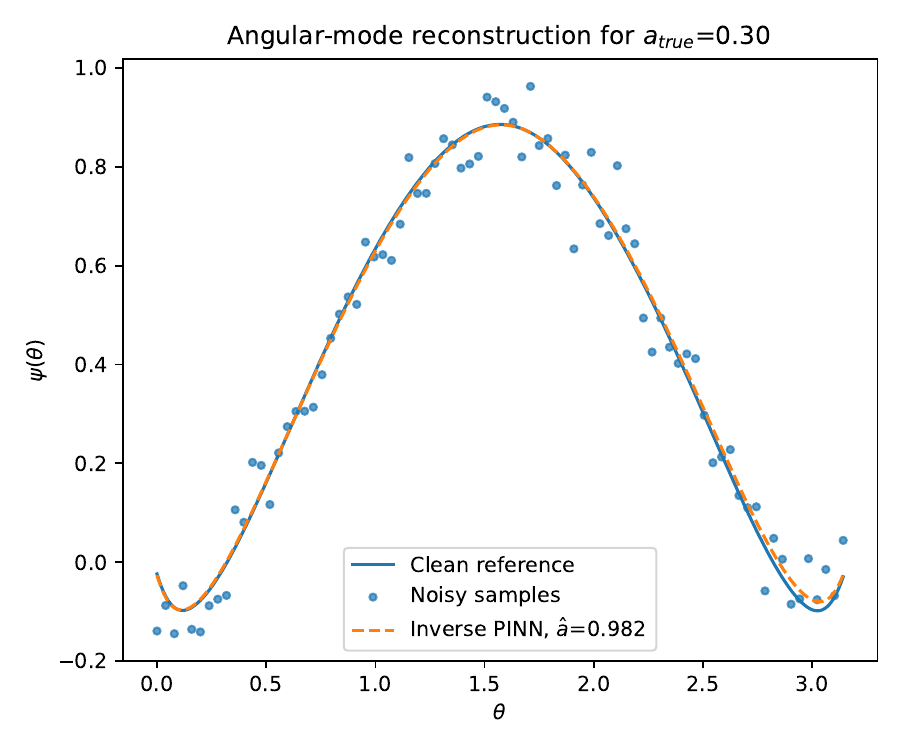}
\caption{$a_{\mathrm{true}}=0.30$}
\end{subfigure}

\begin{subfigure}{0.4\textwidth}
\centering
\includegraphics[width=\textwidth]{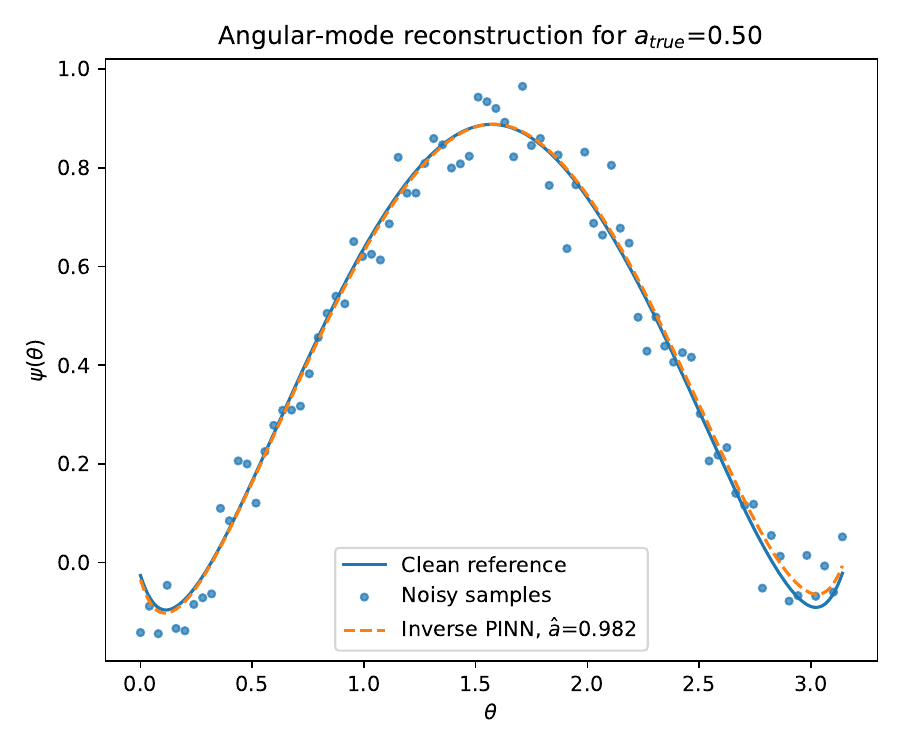}
\caption{$a_{\mathrm{true}}=0.50$}
\end{subfigure}
\hfill
\begin{subfigure}{0.4\textwidth}
\centering
\includegraphics[width=\textwidth]{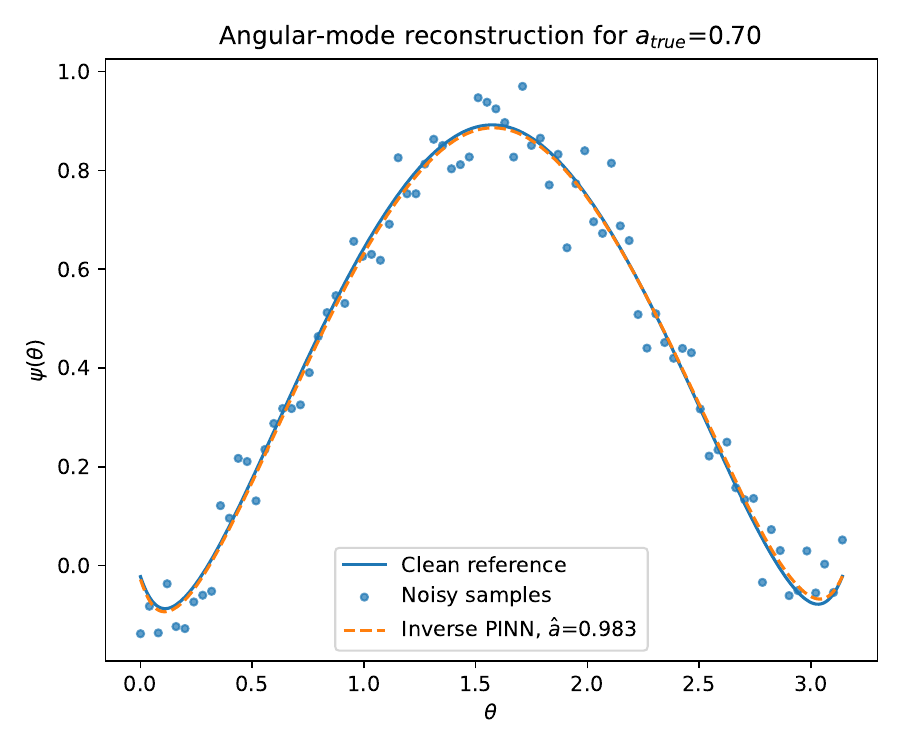}
\caption{$a_{\mathrm{true}}=0.70$}
\end{subfigure}

\begin{subfigure}{0.4\textwidth}
\centering
\includegraphics[width=\textwidth]{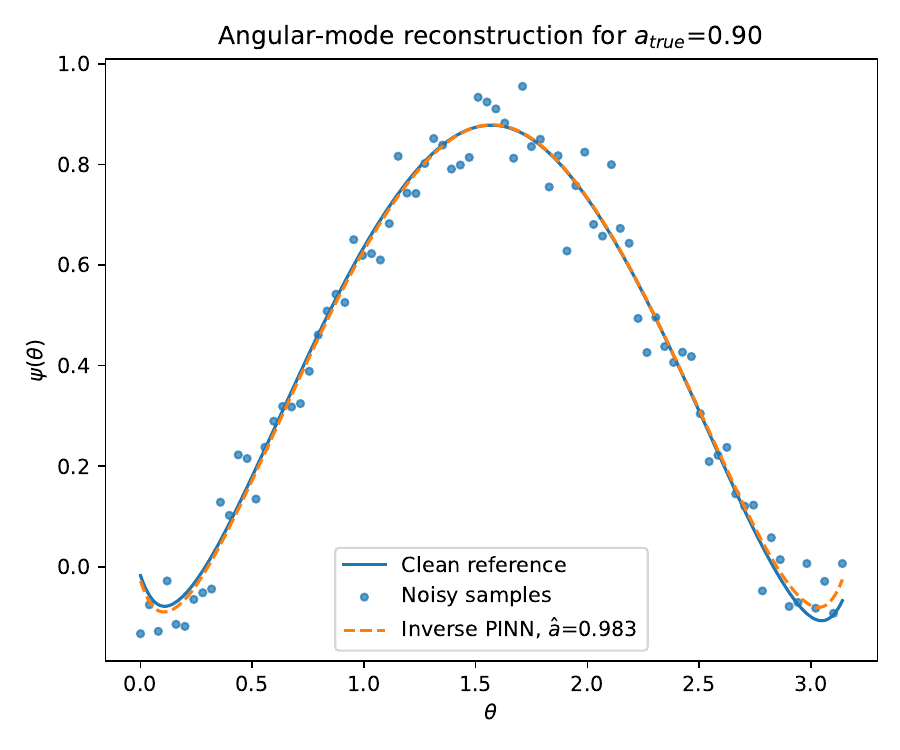}
\caption{$a_{\mathrm{true}}=0.90$}
\end{subfigure}
\caption{Angular-mode reconstruction examples for the supplied synthetic spin-recovery runs. Each panel compares the clean reference profile, noisy samples, and the inverse-PINN reconstruction for the corresponding reference spin.}
\label{fig:solution_profiles}
\end{figure}

\subsection{Identifiability of the Reduced Inverse Spin Problem}
\label{sec:identifiability}

The clean spin-recovery benchmark indicates that the present inverse formulation does not uniquely recover the full range of reference spin values. As shown in Table~\ref{tab:clean_spin_benchmark}, the inferred spin values cluster near the upper part of the allowed spin interval for all tested reference spins. This behavior should not be interpreted as successful spin recovery. Instead, it reveals a weak-identifiability regime of the current reduced inverse problem.

This distinction is central to the interpretation of the benchmark. The inverse PINN is able to reproduce the angular-mode profiles qualitatively, but the recovered spin parameter remains biased. Therefore, accurate angular-profile reconstruction and low residual loss are not sufficient, by themselves, to demonstrate accurate physical-parameter recovery. In inverse physics-informed problems, residual minimization can indicate consistency with the imposed differential operator, but it does not necessarily guarantee that the unknown physical parameter is identifiable from the available data.

A likely explanation is that the reduced scalar angular operator, under the present fixed choices of $m$, $\omega$, and $\lambda$, does not provide sufficiently independent spin-sensitive information to distinguish all reference spin values from angular-mode samples alone. Since the spin enters the residual through the term $\hat{a}^{\,2}\omega^2\cos^2\theta$, changes in $\hat{a}$ may be partially compensated by changes in the learned angular profile $\hat{\psi}_{\Theta}(\theta)$ during optimization. This compensation can lead to qualitatively similar angular reconstructions for different spin values, while the trainable spin parameter is attracted toward a biased region of the allowed interval.

The forward PINN diagnostic and the inverse spin-recovery benchmark therefore answer different questions. The forward diagnostic tests whether the reduced differential operator can be minimized under prescribed physical parameters. The inverse benchmark tests whether the spin parameter itself can be recovered from the available synthetic angular-mode information. The observed clustering of $\hat{a}$ near the upper boundary suggests that the present loss landscape contains a biased attractor for the trainable spin parameter. Consequently, the main value of the benchmark is diagnostic rather than confirmatory: it shows that the current reduced angular-only formulation is not yet sufficient for reliable spin estimation.

These results indicate that additional constraints are required before the framework can be interpreted as a validated spin-estimation pipeline. Possible improvements include alternative loss weighting, explicit identifiability tests, joint inference of the angular separation parameter $\lambda$, richer spin-sensitive data, and extensions beyond the reduced scalar angular setting. In particular, loss-profile scans over fixed spin values, sensitivity analysis of the residual with respect to $a$, and joint $(a,\lambda)$ inference would help determine whether the reduced angular formulation contains enough independent information to separate spin effects from neural-function flexibility.

\subsection{Fixed-Spin Loss-Profile Diagnostic}
\label{sec:loss_profile_scan}

To further diagnose the weak-identifiability behavior observed in the inverse benchmark, a fixed-spin loss-profile scan was performed over the allowed spin interval. In this diagnostic, the spin parameter is not trainable. Instead, it is fixed to a set of candidate values $a_j$, while the neural-network weights are optimized for each candidate spin.

For each reference spin value $a_{\mathrm{true}}$, the profile loss is defined as
\begin{equation}
\mathcal{L}_{\mathrm{profile}}(a_j)
=
\min_{\Theta}
\mathcal{L}_{\mathrm{total}}(\Theta,a_j),
\end{equation}
where $a_j$ denotes a fixed candidate spin value and $\Theta$ denotes the neural-network parameters. If the reduced inverse problem were strongly identifiable, the minimum of $\mathcal{L}_{\mathrm{profile}}(a_j)$ would be expected to occur near the corresponding reference spin $a_{\mathrm{true}}$.

The loss-profile scan therefore provides a direct diagnostic of whether the reduced angular formulation contains enough spin-sensitive information to distinguish different spin values. A shallow or biased loss profile would support the interpretation that the spin parameter is weakly identifiable under the present angular-only formulation.

\begin{figure}[H]
\centering
\includegraphics[width=0.75\textwidth]{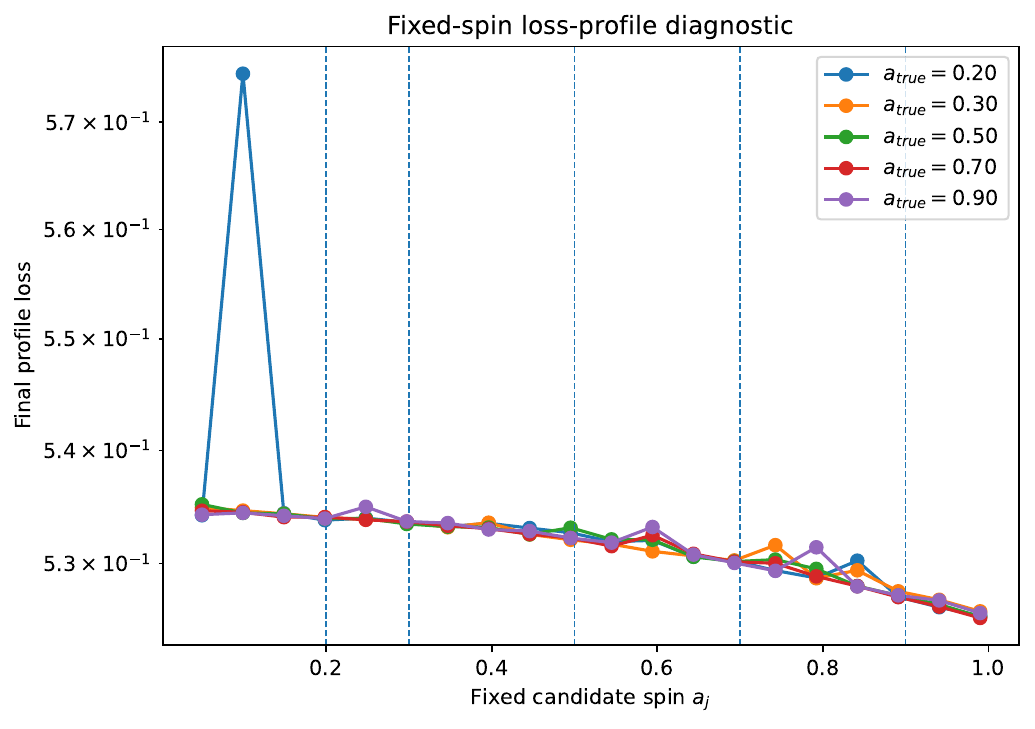}
\caption{Fixed-spin loss-profile diagnostic. For each reference spin value, the spin parameter is fixed across a grid of candidate values and the neural-network weights are optimized. A strongly identifiable formulation would produce profile-loss minima near the corresponding reference spin values.}
\label{fig:loss_profile_scan}
\end{figure}
The resulting loss profiles do not exhibit well-localized minima at the corresponding reference spin values. Instead, the curves remain relatively shallow and show a mild preference toward the upper part of the candidate-spin interval. This behavior is consistent with the direct inverse-PINN results in Table~\ref{tab:clean_spin_benchmark}, where the inferred spin values cluster near the upper part of the allowed interval. The fixed-spin scan therefore supports the interpretation that the present reduced angular-only inverse formulation is weakly identifiable: different candidate spin values can lead to comparable profile losses after optimization of the neural-network weights. The isolated spike observed in one profile curve is likely attributable to stochastic optimization variability in a single fixed-spin training run and does not affect the overall shallow-profile trend.

\subsection{Noise-Contaminated Spin-Recovery Benchmark}

To evaluate robustness, Gaussian perturbations are added to the synthetic angular-mode samples:
\begin{equation}
\psi_{\mathrm{obs}}(\theta_i)
=
\psi_{\mathrm{ref}}(\theta_i)
+
\epsilon_i,
\qquad
\epsilon_i \sim \mathcal{N}(0,\sigma^2).
\end{equation}

The tested noise levels are
\[
\sigma \in \{0.00,0.01,0.03,0.05,0.10\}.
\]

For each noise level, the experiment is repeated across multiple random seeds. The reported quantities are the mean inferred spin, standard deviation of the inferred spin, mean absolute error, and root mean squared error.

\begin{table}[htbp]
\centering
\caption{Noise-contaminated spin-recovery benchmark. Mean and standard deviation of the inferred spin are computed across all reference spin values and random seeds for each Gaussian noise level. MAE and RMSE are computed against the corresponding reference spin values.}
\label{tab:noise_spin_benchmark}
\begin{tabular}{ccccc}
\hline
Noise level $\sigma$ & Mean $\hat{a}$ & Std $\hat{a}$ & MAE & RMSE \\
\hline
0.00 & 0.982457 & 0.000249 & 0.462457 & 0.528626 \\
0.01 & 0.982458 & 0.000239 & 0.462458 & 0.528627 \\
0.03 & 0.982459 & 0.000219 & 0.462459 & 0.528628 \\
0.05 & 0.982460 & 0.000200 & 0.462460 & 0.528629 \\
0.10 & 0.982458 & 0.000164 & 0.462458 & 0.528628 \\
\hline
\end{tabular}
\end{table}

This benchmark directly measures how perturbations in the angular-mode samples affect the inferred spin parameter. It is therefore more informative for spin estimation than residual convergence alone.

\subsection{Methodological Comparison with Baseline Synthetic Models}

Table~\ref{tab:model_comparison} summarizes the methodological role of the proposed inverse PINN relative to simplified baseline synthetic models. The table is not intended as a completed quantitative superiority comparison, because the present study focuses on diagnosing the identifiability behavior of the inverse physics-informed formulation. Instead, the comparison clarifies which models would be needed for a full future benchmark.

\begin{table}[htbp]
\centering
\caption{Methodological comparison between synthetic baseline models and the proposed inverse PINN. The present manuscript quantitatively evaluates the proposed inverse PINN and treats the remaining models as future controlled comparators.}
\label{tab:model_comparison}
\begin{tabular}{lccc}
\hline
Model & Physics residual & Spin treated as unknown & Role in this study \\
\hline
Data-only neural regression & No & Yes & Future comparator \\
Least-squares synthetic fit & Partial & Yes & Future comparator \\
Forward PINN diagnostic & Yes & No & Residual-convergence check \\
Proposed inverse PINN & Yes & Yes & Quantitatively evaluated \\
\hline
\end{tabular}
\end{table}

The forward PINN diagnostic is included to show residual minimization for prescribed spin values, but it is not a spin-estimation model because the spin parameter is fixed before training. The relevant inverse problem is whether the spin parameter can be recovered from angular-mode information. Therefore, the present comparison should be interpreted as a methodological positioning table rather than as evidence of numerical superiority over all possible synthetic baselines.

\subsection{True vs Estimated Spin}

\begin{figure}[htbp]
\centering
\includegraphics[width=0.75\textwidth]{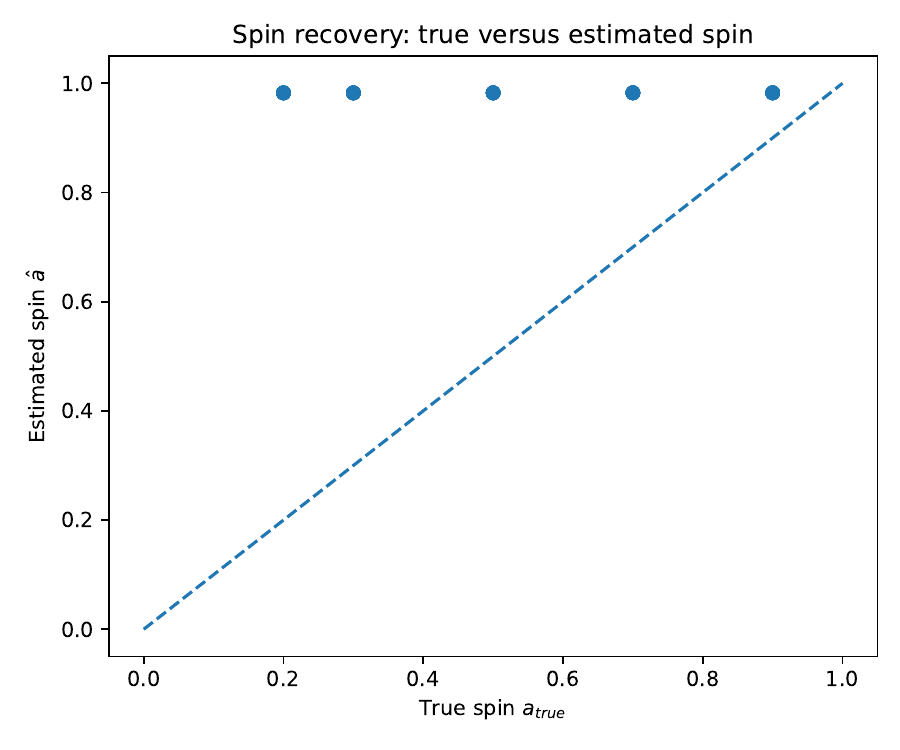}
\caption{True spin versus estimated spin for the inverse physics-informed framework. The diagonal line indicates ideal recovery.}
\label{fig:true_vs_estimated_spin}
\end{figure}

The relationship between the reference spin values and the inferred spin values is shown in Figure~\ref{fig:true_vs_estimated_spin}. The diagonal line corresponds to ideal recovery, $\hat{a}=a_{\mathrm{true}}$.

In the supplied runs, the inferred spin values cluster near the upper part of the allowed interval rather than following the diagonal reference line. This indicates that, under the current configuration, the inverse optimization does not yet recover the full range of reference spins accurately. The deviation from the diagonal is therefore interpreted as a diagnostic of the present formulation and motivates additional constraints, reweighting, or identifiability checks in future experiments.

\subsection{Noise Sensitivity of Spin Recovery}

\begin{figure}[htbp]
\centering
\includegraphics[width=0.75\textwidth]{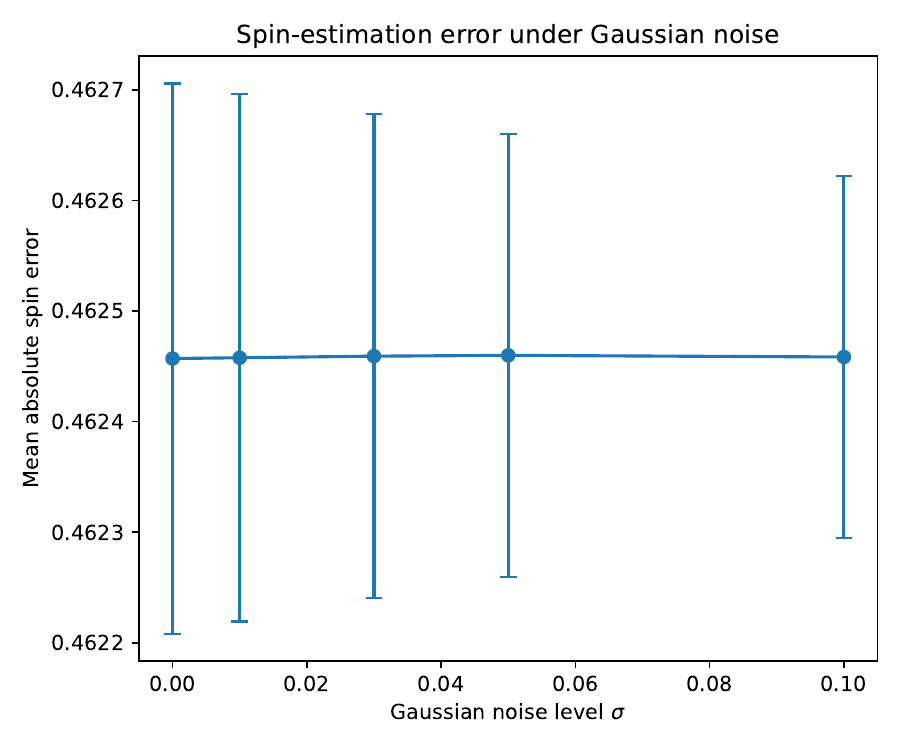}
\caption{Mean absolute spin-estimation error as a function of Gaussian noise amplitude. Error bars indicate variability across random seeds.}
\label{fig:noise_vs_spin_error}
\end{figure}

Figure~\ref{fig:noise_vs_spin_error} shows the spin-estimation error as a function of the Gaussian noise amplitude. Error bars indicate variability across random seeds.

This figure evaluates whether the inferred spin remains stable as the angular-mode samples are increasingly perturbed. In the supplied runs, the mean absolute spin-estimation error remains nearly unchanged across the tested Gaussian noise amplitudes, which suggests that the dominant source of error is the bias of the inverse spin estimate rather than the added synthetic noise. The purpose is not to reproduce a specific detector noise model, but to assess the sensitivity of the inverse formulation to controlled perturbations in the input angular-mode information.

\section{Discussion and Limitations}
The numerical results in Sections~4 and~5 show that the trained PINN can reproduce the qualitative angular-mode profiles in the reduced synthetic setting while maintaining low residual errors. However, the inverse results also show that qualitative profile reconstruction and residual minimization are not sufficient to guarantee accurate spin-parameter recovery. These findings are in agreement with previous studies such as \cite{luna2023solving}.
Several methodological merits of the PINN-based formulation can be identified:
\begin{itemize}
    \item The physics-informed residual constrains the learned angular function to remain close to the solution manifold of the reduced differential operator, improving interpretability relative to unconstrained neural regression.
    \item The trainable spin parameter is inferred within the same optimization problem as the angular mode function, allowing the inverse problem to be evaluated directly through parameter-recovery metrics.
    \item The separation between forward residual convergence and inverse spin recovery clarifies that a small PDE residual alone is not sufficient evidence of accurate physical-parameter inference.
\end{itemize}
However, several limitations should be considered in this study. First, the Teukolsky equation was reduced to a scalar, 1D form to allow experimentation and visualization. Thus, the complete perturbed gravitational fields in Kerr spacetime, in particular in the case of two-spin fields, were not addressed. Second, the current PINN implementation remains computationally more expensive than traditional solvers for low-dimensional systems due to the cost of automatic differentiation and iterative training.

Lastly, although convergence is attained for a moderate spin value ($a = 0.7$), the performance degradation was near extremal values ($a \to 1$),  in accordance with the results of \cite{cornell2024solving}. More architectural changes or sampling strategies may be required to address these regimes.

\subsection{Interpretation of the Benchmark}

The benchmark results should be interpreted within the reduced synthetic setting used in this work. The clean and noise-contaminated tests evaluate whether the proposed inverse PINN can recover the spin parameter from angular-mode samples when the reference spin is known. This is different from a full observational black-hole spin-estimation pipeline, where additional effects such as detector response, source inclination, accretion geometry, calibration uncertainties, and model degeneracies must be included.

The main purpose of the benchmark is therefore methodological. It tests the behavior of the inverse physics-informed formulation under controlled synthetic conditions and separates residual convergence from direct spin-parameter recovery. Because the supplied baseline result files contain only the inverse-PINN runs, the present comparison should be interpreted as a diagnostic benchmark rather than as a completed quantitative superiority claim over data-only or least-squares alternatives.

However, the benchmark does not by itself establish observational accuracy. Validation against high-resolution Kerr perturbation solvers and astrophysical data remains necessary before the method can be considered a fully operational spin-estimation pipeline.

\subsection{Spin-Bias Interpretation}
\label{sec:spin_bias_interpretation}

The most important outcome of the inverse benchmark is the systematic clustering of the inferred spin near the upper part of the allowed interval. This behavior shows that the present reduced inverse formulation is stable across seeds and noise levels, but stable toward a biased solution. Therefore, robustness should not be confused with accuracy. The nearly unchanged MAE across Gaussian noise levels indicates that the dominant error source is not the added random perturbation, but the structure of the inverse optimization problem itself.

This result has two implications. First, it shows that angular-profile reconstruction and residual minimization are insufficient by themselves to validate spin inference. A model may reproduce the angular profile qualitatively while still failing to recover the physical spin parameter. Second, it motivates the use of additional diagnostic tools, including loss-ablation experiments, spin-wise residual scans, joint eigenvalue inference, and comparison with numerical Kerr perturbation solvers. In this sense, the present study provides a diagnostic framework for identifying where reduced physics-informed spin inference succeeds and where it remains underconstrained.

\section{Conclusion}

This work presented a diagnostic hybrid physics-informed framework for synthetic black-hole spin inference using a reduced scalar angular Teukolsky-like equation. The proposed formulation treats spin recovery as an inverse problem: the angular mode function is approximated by a Physics-Informed Neural Network, while the spin parameter is optimized as a trainable physical quantity constrained by the reduced differential operator, angular-mode data consistency, boundary conditions, normalization, and regularization.
The study separates forward residual convergence from inverse spin recovery. The forward diagnostic confirms that the reduced angular operator can be minimized under prescribed spin parameters. However, the inverse benchmark shows that the current formulation does not accurately recover the full range of reference spins. Instead, the inferred values cluster near the upper part of the allowed spin interval. This behavior is interpreted as an identifiability and loss-geometry limitation of the present reduced inverse problem rather than as a validated astrophysical spin-estimation result.
The main contribution is therefore methodological and diagnostic. The framework provides a controlled environment for testing whether physics-informed constraints can support spin-sensitive inverse modeling, while also revealing when angular-profile reconstruction and residual minimization are insufficient for reliable physical-parameter recovery. The observed spin bias motivates further work on loss reweighting, identifiability analysis, adaptive sampling, joint inference of the angular separation eigenvalue, and richer uncertainty models.
The current formulation remains limited to a reduced scalar angular setting and synthetic angular-mode data. Future work should extend the framework toward the full Teukolsky equation, complex mode frequencies, spin-weighted angular functions, realistic observational uncertainty models, and validation against high-resolution Kerr perturbation solvers and astrophysical datasets.

\section*{Data availability}

The code and supporting material are publicly available at:
\url{https://github.com/stmenzi/hybrid-physics-informed-black-hole-spin-estimation}.

\section*{Funding}

The author received no external funding for this research.

\section*{Competing interests}

The author declares no competing interests.

\section*{Declaration of AI-assisted language editing}

During the preparation of this manuscript, the author used an AI-based tool for language refinement, clarity improvement, and coherence editing. The tool was not used to generate scientific results, perform data analysis, or replace the author’s scientific interpretation. The author reviewed, edited, and approved the final manuscript and takes full responsibility for its content.

\vspace{0.5cm}

\bibliographystyle{plain}
\bibliography{sample}

@inproceedings{mcclintock2011measuring,
  title={Measuring the spins of stellar-mass black holes},
  author={McClintock, Jeffrey},
  booktitle={APS April Meeting Abstracts},
  volume={2011},
  pages={R3--001},
  year={2011}
}

@article{reynolds2014measuring,
  title={Measuring black hole spin using X-ray reflection spectroscopy},
  author={Reynolds, Christopher S},
  journal={Space Science Reviews},
  volume={183},
  pages={277--294},
  year={2014},
  publisher={Springer}
}

@article{garcia2014improved,
  title={Improved reflection models of black hole accretion disks: treating the angular distribution of X-rays},
  author={Garc{\'\i}a, J and Dauser, T and Lohfink, A and Kallman, TR and Steiner, JF and McClintock, JE and Brenneman, L and Wilms, J and Eikmann, W and Reynolds, CS and others},
  journal={The Astrophysical Journal},
  volume={782},
  number={2},
  pages={76},
  year={2014},
  publisher={IOP Publishing}
}

@article{motta2015geometrical,
  title={Geometrical constraints on the origin of timing signals from black holes},
  author={Motta, SE and Casella, Piergiorgio and Henze, M and Mu{\~n}oz-Darias, T and Sanna, A and Fender, R and Belloni, T},
  journal={Monthly Notices of the Royal Astronomical Society},
  volume={447},
  number={2},
  pages={2059--2072},
  year={2015},
  publisher={Oxford University Press}
}

@article{stuchlik2016models,
  title={Models of quasi-periodic oscillations related to mass and spin of the GRO J1655-40 black hole},
  author={Stuchl{\'\i}k, Zden{\v{e}}k and Kolo{\v{s}}, Martin},
  journal={Astronomy \& Astrophysics},
  volume={586},
  pages={A130},
  year={2016},
  publisher={EDP Sciences}
}

@article{luna2023solving,
  title={Solving the Teukolsky equation with physics-informed neural networks},
  author={Luna, Raimon and Calder{\'o}n Bustillo, Juan and Seoane Mart{\'\i}nez, Juan Jos{\'e} and Torres-Forn{\'e}, Alejandro and Font, Jos{\'e} A},
  journal={Physical Review D},
  volume={107},
  number={6},
  pages={064025},
  year={2023},
  publisher={APS}
}

@article{cornell2024solving,
  title={Solving the Regge-Wheeler and Teukolsky equations: supervised versus unsupervised physics-informed neural networks},
  author={Cornell, Alan S and Herbst, Sheldon R and Ncube, Anele M and Noshad, Hajar},
  journal={arXiv preprint arXiv:2402.11343},
  year={2024}
}

@article{leaver1985analytic,
  title={An analytic representation for the quasi-normal modes of Kerr black holes},
  author={Leaver, Edward W},
  journal={Proceedings of the Royal Society of London. A. Mathematical and Physical Sciences},
  volume={402},
  number={1823},
  pages={285--298},
  year={1985},
  publisher={The Royal Society London}
}

@article{berti2006gravitational,
  title={Gravitational-wave spectroscopy of massive black holes with the space interferometer LISA},
  author={Berti, Emanuele and Cardoso, Vitor and Will, Clifford M},
  journal={Physical Review D—Particles, Fields, Gravitation, and Cosmology},
  volume={73},
  number={6},
  pages={064030},
  year={2006},
  publisher={APS}
}

@article{baron2019machine,
  title={Machine learning in astronomy: A practical overview},
  author={Baron, Dalya},
  journal={arXiv preprint arXiv:1904.07248},
  year={2019}
}

@article{raissi2019physics,
  title={Physics-informed neural networks: A deep learning framework for solving forward and inverse problems involving nonlinear partial differential equations},
  author={Raissi, Maziar and Perdikaris, Paris and Karniadakis, George E},
  journal={Journal of Computational physics},
  volume={378},
  pages={686--707},
  year={2019},
  publisher={Elsevier}
}

@article{Menziltsidou2026EPJP,
  author  = {Menziltsidou, Stella},
  title   = {A physics-informed neural network solver for a simplified scalar angular Teukolsky-like equation},
  journal = {The European Physical Journal Plus},
  year    = {2026},
  doi     = {10.1140/epjp/s13360-026-08019-3}
}

@article{Yang2019AUQ,
  author  = {Yang, Yibo and Perdikaris, Paris},
  title   = {Adversarial uncertainty quantification in physics-informed neural networks},
  journal = {Journal of Computational Physics},
  volume  = {394},
  pages   = {136--152},
  year    = {2019},
  doi     = {10.1016/j.jcp.2019.05.027}
}

@article{Yang2021BPINN,
  author  = {Yang, Liu and Meng, Xuhui and Karniadakis, George Em},
  title   = {B-PINNs: Bayesian physics-informed neural networks for forward and inverse PDE problems with noisy data},
  journal = {Journal of Computational Physics},
  volume  = {425},
  pages   = {109913},
  year    = {2021},
  doi     = {10.1016/j.jcp.2020.109913}
}

@article{Teukolsky1973,
  author  = {Teukolsky, Saul A.},
  title   = {Perturbations of a Rotating Black Hole. I. Fundamental Equations for Gravitational, Electromagnetic, and Neutrino-Field Perturbations},
  journal = {The Astrophysical Journal},
  volume  = {185},
  pages   = {635--647},
  year    = {1973},
  doi     = {10.1086/152444}
}

@article{ReggeWheeler1957,
  author  = {Regge, Tullio and Wheeler, John A.},
  title   = {Stability of a Schwarzschild Singularity},
  journal = {Physical Review},
  volume  = {108},
  number  = {4},
  pages   = {1063--1069},
  year    = {1957},
  doi     = {10.1103/PhysRev.108.1063}
}

@article{Zerilli1970,
  author  = {Zerilli, Frank J.},
  title   = {Effective Potential for Even-Parity Regge-Wheeler Gravitational Perturbation Equations},
  journal = {Physical Review Letters},
  volume  = {24},
  number  = {13},
  pages   = {737--738},
  year    = {1970},
  doi     = {10.1103/PhysRevLett.24.737}
}
\end{document}